\documentclass[twocolumn,showpacs,preprintnumbers,amsmath,amssymb]{revtex4}
\usepackage[dvips]{graphicx}
\usepackage{graphics}
\usepackage{epsf}
\begin{document}

\newcommand{\e}[1]{\times 10^{#1}}

\newlength{\figwidtha}
\newlength{\figwidthb}
\setlength{\figwidtha}{0.85\columnwidth} 
\setlength{\figwidthb}{0.76\columnwidth}

\title{Kinetic Monte Carlo simulation of shape transition of strained quantum dots}


\author{Chi-Hang Lam}
\affiliation{
Department of Applied Physics, Hong Kong Polytechnic University,
Hung Hom, Hong Kong, China
}

\date{\today}

\begin{abstract}
  The pyramid-to-dome transition in Ge$_{x}$Si$_{1-x}$ on Si(100)
  initiated by step formation on pyramidal quantum dots is
  atomistically simulated using a multistate lattice model in
  two-dimensions incorporating effective surface
  reconstructions. Under quasi-equilibrium growth
  conditions associated with low deposition rates, the transition
  occurs at island size $n_c$ following $\sqrt{n_c}
  \sim x^{-1.69}$ independent of temperature and deposition rate.
  The shape transition is found to be an activated process.
  Results are explained by a theory based on simple forms of facet
  energies and elastic energies estimated using a shallow island
  approximation. An asymptotic scaling relation $n_c^{1/d} \sim
  x^{-2}$ for $x\rightarrow 0$ applicable to $d$ = 2 or 3 dimensions
  is derived. The shape transition energy barrier can be 
  dominated by the interface energy between steep and shallow facets.  
\end{abstract}


\maketitle

\section{introduction}

The self assembly of quantum dots in heteroepitaxy exhibits very
interesting physics and has possible application to device fabrication
\cite{Politi2000,Shchukin2003,Berbezier2009}. Growth of Ge or GeSi
alloy nanostructures on Si(100) is the prototype example most widely
studied.
Under typical deposition conditions, pyramidal islands bounded by shallow
(105) facets form spontaneously on a wetting layer. They
can then undergo transitions into multi-faceted dome islands dominated by much
steeper (113) facets and bounded also by other facets \cite{Ross1999}.  This shape transition gives
rise to a bimodal island size distribution with enhanced dome size
uniformity \cite{Ross1998}. An atomic pathway based on step bunching
on the pyramids has been identified
\cite{Montalenti2004,Montalenti2007}.

In this work, we report atomistic dynamic simulations of the
pyramid-to-dome transition using a fast kinetic Monte Carlo (KMC)
approach based on a multistate solid-on-solid model in two-dimensions
(2D). Extensive simulations under a wide range of conditions are
performed and a simple analytical description is presented. A scaling
relation for the transition island size generalizable to three
dimensions (3D) is investigated.

Kinetic simulation of a strained film is much more challenging computationally
than the unstrained case because of the long-range nature of elastic
interactions.  First principles calculations
\cite{Fujikawa2002,Montalenti2005,Lu2005,Shklyaev2005,Montalenti2007}
and molecular-dynamics simulations \cite{Maxim2003,Retford2007} have provided
important information on the energetics of the relevant surfaces and
steps, but are in general limited to the studies of static properties
at small system sizes. Continuum simulations in contrast are
instrumental for investigating large scale and long time
behaviors \cite{Tu2004,Chiu2007}, but lacking atomic discreteness,
nucleation events associated with island formation and shape
transition cannot be naturally simulated. KMC simulations based on
lattice models are hence unique in allowing large scale atomistic
studies on the dynamics of strained heteroepitaxy.

Using a ball and spring lattice model for elastic solids, Orr et. al
\cite{Orr1992} conducted early KMC simulations of strained layers in
2D. In the simulations, the elastic
energy of the system has to be computed repeatedly in order
to simulate the atomic hopping events responsible for the
morphological evolution.
Applying more advanced algorithms for the solution of the elastic
problem and the sampling of atomic hopping events, simulations using
large and moderate system sizes in 2D
\cite{Lam2002,Gray2005,Schulze2009} and 3D
\cite{Lung2005,Smereka2006,Lam2007,Lam2008,Smereka2009} respectively
became possible. The model was extended recently to model
(105) facets \cite{Lam2010} and atomic intermixing with substrate
atoms \cite{Smereka2010}.  Alternatively, KMC simulations can also be
performed efficiently using more approximate forms of elastic
interactions \cite{Ratsch1996,Meixner2001,Zhu2007}.

This paper is organized as follows. Section II explains our multistate
model for elastic solids which can account for both a shallow and a
steep facet. The KMC simulation results are presented in Sec. III. In
Sec. IV, island energies and the island transition rate are calculated
theoretically. In Sec. VI, a scaling relation between the transition
island size and the Ge concentration is derived. We conclude in
Sec. VI with some further discussions.

\section{multistate solid on solid model}

Our model is based on a ball and spring square lattice model of elastic
solids for Ge$_x$Si$_{1-x}$ on Si \cite{Lam2002}.
The substrate lattice constant is $a_s=2.72$ \AA~ while the film material admits a lattice misfit $\epsilon = 0.04 x$.  Each node on the lattice represents an atom and it is connected to its nearest and next nearest neighbors by elastic springs with force constants $k_N=13.85$~eV/$a_s^2$ and $k_{NN}=k_N/2$ respectively.  This choice gives the correct modulus $c_{11}$ of silicon and a shear modulus constant along tangential and diagonal directions.

\begin{figure}[htp]
{\epsfxsize 0.75\figwidtha \epsfbox{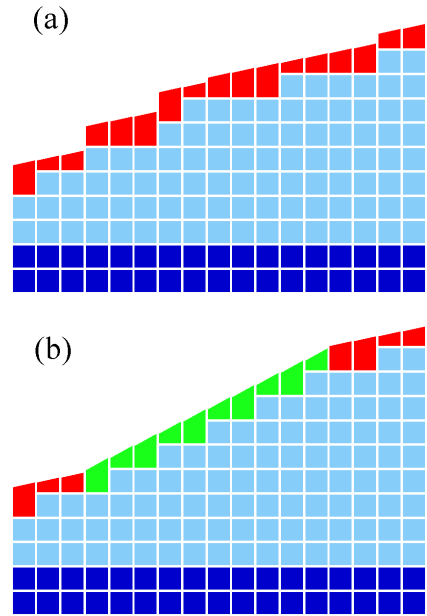}}
\caption{
  \label{F:surf}
  (a) A shallow facet with steps leading to (b) a steep facet
  from a
  small-scale simulation. In (b), the first 6 surface atoms from the left
  have local deformation states $(\sigma_i,
  \kappa_i)=(\frac15,\frac25),(\frac15,-\frac25),
  (\frac15,-\frac15),(\frac12,\frac14),(\frac12,-\frac14)$ and
  $(\frac12,\frac14)$. Surface atoms in shallow (steep) facets are
  shaded in red (green), while bulk Ge (Si) atoms are shaded in light
  (dark) blue.
}
\end{figure}

In this work, (100), (105) and (113) surfaces must 
be effectively simulated.  A leveled surface in the model naturally
accounts for a (100) surface. However, lattice models generally lead
to islands with a single type of sidewalls at 45$^\circ$ inclination
or steeper
\cite{Orr1992,Lam2002,Gray2005,Schulze2009,Lung2005,Smereka2006,Lam2007,Lam2008,Smereka2009}.
A multi-state extension has been introduced recently
in Ref. \cite{Lam2010} to effectively model the much shallower (105) facets
of a pyramid in 2D. We now
further generalize it to simulate both shallow and steep facets with slopes 
\begin{equation}
  \label{eq:s}
s_1=1/5 \mbox{~~~ and ~~~~} s_2=1/2
\end{equation}
comparable to those of realistic
facets in pyramids and domes.
Specifically, atoms are normally represented by squares. To effectively model
surface reconstructions leading to specific facets, we allow local
deformation of the topmost atoms in the film or substrate into
trapezoids each characterized by a tilt variable $\sigma_i$ and an
extension variable $\kappa_i$.  Here, $\sigma_i$ gives the slope of
the upper surface of an atom and equals
\begin{equation}
  \sigma_i = 0,  ~ \pm 1/5 , \mbox{~ or ~} ~\pm 1/2 
\end{equation}
at a locally undeformed region, a shallow facet or a steep facet respectively.
Allowing atomistically flat shallow and steep facets further requires
additional freedoms of vertical stretching or compression of the topmost
atoms by
\begin{equation}
  \kappa_i = 
    \left\{ 
        \begin{array}{ll}
            0 & \mbox{for $\sigma_i = 0$}\\
        0 ,  ~ \pm 1/5,  \mbox{~ or ~} \pm 2/5 ~~~ & \mbox{for $\mid \sigma_i\mid=1/5$}\\
        \pm   1/4 ~~~ & \mbox{for $\mid\sigma_i\mid= 1/2$}
        \end{array}
    \right. 
\end{equation}
This characterizes a total of $15$ possible local deformation
states. 
All lengths are measured in unit of $a_s$ throughout this paper.
Atomic
column $i$ with $h_i$ atoms is thus trapezoidal in general with the
left and right edges of heights $h_i^a$ and $h_i^b$ given by
\begin{eqnarray}
  h_i^a &=& h_i + \kappa_i - \frac{\sigma_i}2\\
  h_i^b &=& h_i + \kappa_i + \frac{\sigma_i}2.
\end{eqnarray}
A surface step in between column $i$ and $i+1$
has a height 
\begin{equation}
\delta_i = \mid h_{i+1}^a - h_{i}^b \mid
\end{equation}
projected along the vertical direction.
Figures \ref{F:surf}(a)-(b) show examples of atomic configurations. 

\newcommand{\Dh}{\delta_i} Misfit induced elastic strain is assumed to
be completely independent of the local deformations associated
with faceting introduced above.  The elastic relaxation energy $E_s$ of the system is
defined as the total energy storied in all springs at mechanical
equilibrium compared with that in the homogeneously strained state.
The bond energy of the system is defined relative to that of a flat
surface by
\begin{eqnarray}
\label{Eb}
E_b  &=& \sum_i \left[ 
\phi_{\alpha_i} +  \psi(i,i+1) 
+\omega_{\alpha_i \alpha_{i+1}}( \Dh )
\right]+ E_R ~
\end{eqnarray}
where the facet-type label $\alpha_i$, depending on
$\mid\sigma_i\mid$, indicates if column $i$ is locally undeformed 
($\alpha_i=0$) or corresponds to a shallow ($\alpha_i=1$) or steep facet
($\alpha_i=2$).  The facet formation energy per site
$\phi_{\alpha_i}$ equals $\phi_0=0$, $\phi_1 = 5$~meV, or
$\phi_2=50$~meV. 
These values control the relative stability of the facets in our simulations. They are chosen empirically so that shallow and steep facets start to emerge on islands of appropriate sizes.
Also, the facet interface energy $\psi(i,i+1)$ is
non-zero only at the boundary between either different facet types or
different facet orientations
(i.e. $\sigma_i \neq \sigma_{i+1}$) 
where it equals
$\psi_{\alpha_i\alpha_{i+1}}$ with $\psi_{01}=\psi_{11}=0.35$~eV,
$\psi_{12}=\psi_{22}=0.5$~eV, and $\psi_{02}=\psi_{01}+\psi_{12}$,
assuming $\psi_{\alpha\alpha'}=\psi_{\alpha'\alpha}$.
The $\omega$ term represents surface step energy. It equals
$\gamma\delta_i/2$ on a locally undeformed region where $\gamma=0.5$~eV is the
nearest neighboring bond energy.
If the site $i$ or $i+1$ belongs to a shallow or a steep facet,
it equals ${\beta_\alpha} [ 
    1+\chi - \chi \exp({ 1 - { \Dh/s_\alpha }}) ] +
    {\gamma} (  \Dh   - s_\alpha) /2$
where $\alpha$ is the larger of
$\alpha_i$ and $\alpha_{i+1}$.
Here, $\beta_1=0.3$~eV and $\beta_2=0.2$~eV represent 
the single step energies on shallow and steep facets respectively and
$\chi=0.3$ dictates the tendency of step bunching. 
To discourage very steep regions, the intrinsic step height defined by
$\delta_i' = \mid h_{i+1}-h_i\mid$ disregarding local deformation is
constrained to $\delta_i'\le 1$ and furthermore each pair of
consecutive upward or downward intrinsic steps with $\delta_i'=1$
contributes 0.15~eV to the repulsion energy $E_R$. The constraint also limits
the step heights $\delta_i$ to bounded values, although double steps
in particular, which have
heights $2/5$ and $1$ on shallow and steep facets respectively are
still possible.


The KMC approach simulates every hopping event of a topmost film atom $m$
according to the rate 
\begin{equation}
\label{rate2}
\Gamma(m) = 
{R_0}\exp \left[ \frac{\Delta E_b(m)  + \Delta E_s(m) +
    E_0'}{k_{B}T}\right].
\end{equation}
Atoms then lands random on any other site at most 8 columns away.
Here, $\Delta E_b(m)$ and $\Delta E_s(m)$ denote the change in the
bond energy $E_b$ and the strain energy $E_s$ of the system when the
site is occupied versus unoccupied. 
We put $E_o'=-\gamma-0.67$~eV
and $R_o = 4.1 \times 10^{11} s^{-1}$. This gives the appropriate
adatom diffusion coefficient for silicon (100).
Due to the long-range nature of elastic
interactions, the repeated calculations of $\Delta E_s(m)$ dominates
the simulation run time and we
handle it using a Green's function method together with a
super-particle approach \cite{Lam2008}.
Exposed substrate atoms are not allowed to hop.
Elastic couplings of adatoms with
the rest of the system are weak and are neglected for better
computational efficiency.
Atomic hoppings are assumed to preserve the local deformation states.
After every period $\tau$, a set of deformation states will be
updated.  We put $\tau = 2/\Gamma_{ad}$ where $\Gamma_{ad}$ is the
adatom hopping rate on an locally undeformed region easily calculable
from Eq. (\ref{rate2}).  At an odd (even) numbered updating event, all odd
(even) lattice sites will be updated. The variables $\sigma_i$ and
$\kappa_i$ at those sites are re-sampled from the
allowed set of 15 possible combinations using a heat bath algorithm
based on the relative probability $\exp(-E_{b}/kT)$.
Our model obeys detailed balance.
The dynamical rules described above reduces back to those used in
Refs.  \cite{Lam2002} at locally undeformed regions. 
%
We have critically
checked our software implementation, in particular using a Boltzmann's
distribution test \cite{Lam2008}, which is found to be indispensable
in verifying that practically all, but not only the dominating hopping pathways
can be correctly simulated. We also have checked in small scale
simulations that restricting hoppings to only nearest neighboring
sites rather than allowing long jumps gives similar results except for
an insignificant shift in the time scale.  Wetting layers on the
substrates are believed to be relatively immobile and are not simulated
for simplicity.

\section{Simulation results}

\begin{figure}[htp]
{\epsfxsize 0.79\figwidtha \epsfbox{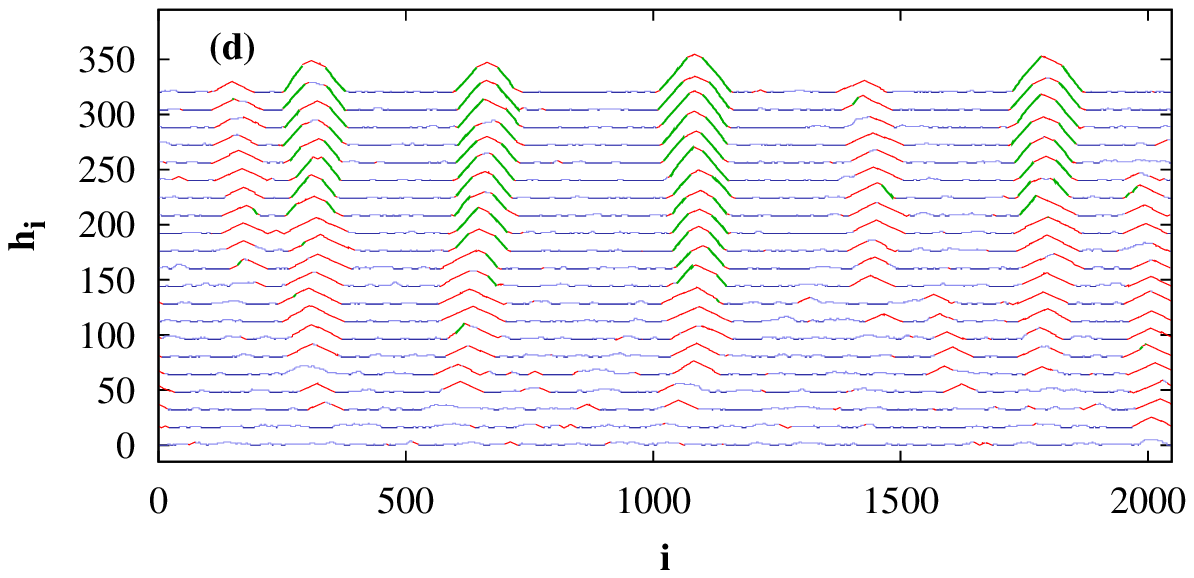}}
{\epsfxsize 0.6\figwidtha \epsfbox{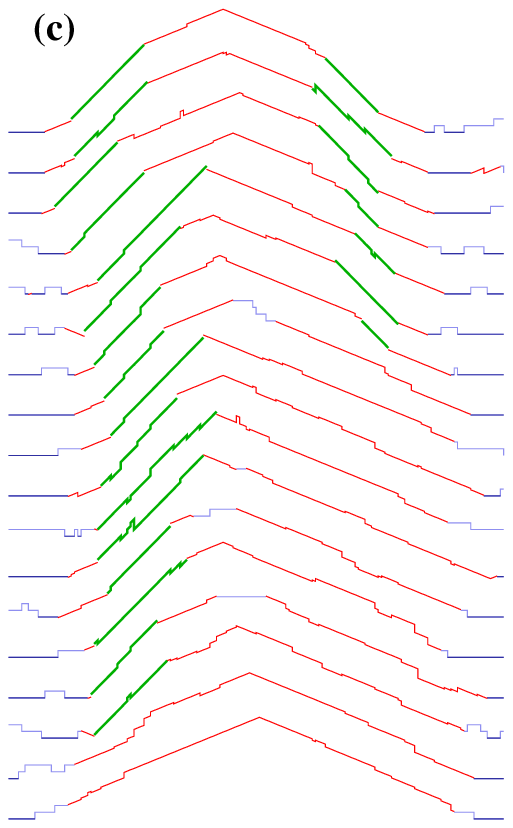}}
\caption{
  \label{F:profile}
  (a)~Surface profiles at 1 to 6 ML coverage simulated at 700$^\circ$C
   with deposition rate 4 ML/s on a lattice of width 2048.  (b) Detailed profiles at 4 to
  4.5 ML coverage showing a
  pyramid-to-dome transition corresponding to the leftmost dome in
  (a).  In (a) and (b), each successive surface corresponds to the deposition of a
  further 1/4 and 1/64 layers respectively and is
  displaced vertically for clarity. 
  Locally deformed regions, shallow
  facets and steep facets are shaded in blue, red and green respectively.
  }
\end{figure}

Figures \ref{F:surf}(a) show a shallow facet with steps from a small
scale simulation. The steps are subsequently smoothed out by the
formation of a steep facet as shown in Fig. \ref{F:surf}(b). 
Figure \ref{F:profile}(a) shows snapshots of a surface from a typical
simulation of deposition at 4 ML/s, 700$^\circ$C and $x=1$ on a
substrate of width 2048. Successive profiles are displaced vertically
for clarity.  Stepped mounds first develop and some of them matures
into pyramids bounded by shallow facets as explained in
Ref. \cite{Lam2010}. Some of the pyramids further turns into domes
bounded mainly by steep facets often with regions of shallow facets at
the top. Figure \ref{F:profile}(b) shows the detailed evolution of one
of the domes.  Steep facets on either side of a pyramid form
independently. The transition hence often goes through a meta-stable
half-dome stage.
The formation of most domes is preceded by steps appearing on the
shallow facets as shown in Fig. \ref{F:surf} and
Fig. \ref{F:profile}(b) and as proposed in Ref. \cite{Montalenti2004}.
A close examination reveal that these steps are highly dynamic and
continuously bunches, separate and diffuses around. After accumulating
a considerable total step height, they transform highly reversibly
into a steep facet. As the total height of the steps increases, the
resulting steep facet is more stable and eventually become fully
stabilized.  A smaller number of domes are initiated instead by the
formation of steep facets at the base of the pyramid when shallow
facets temporarily decay into unfaceted regions due to thermal excitations.

\begin{figure}[htp]
\centerline{\epsfxsize 0.99\figwidthb \epsfbox{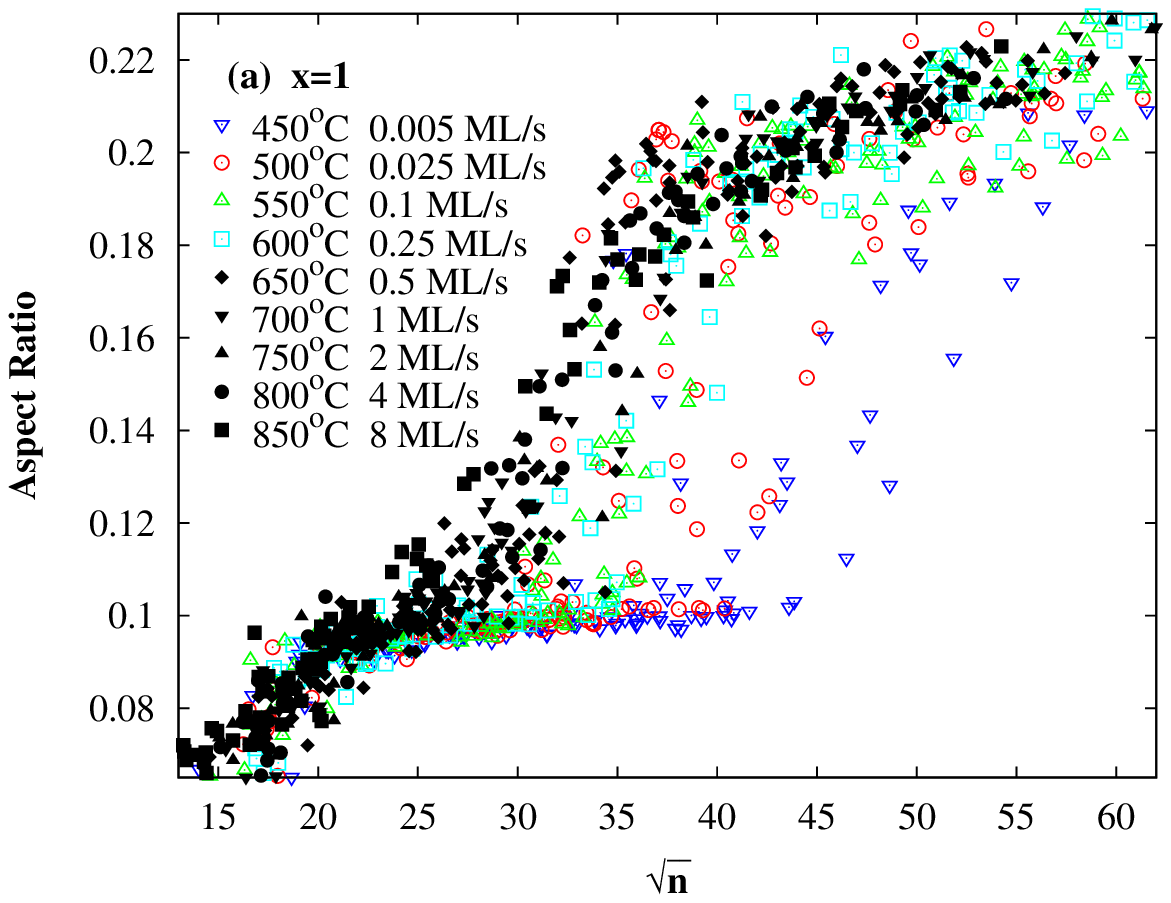}}
\centerline{\epsfxsize 0.99\figwidthb \epsfbox{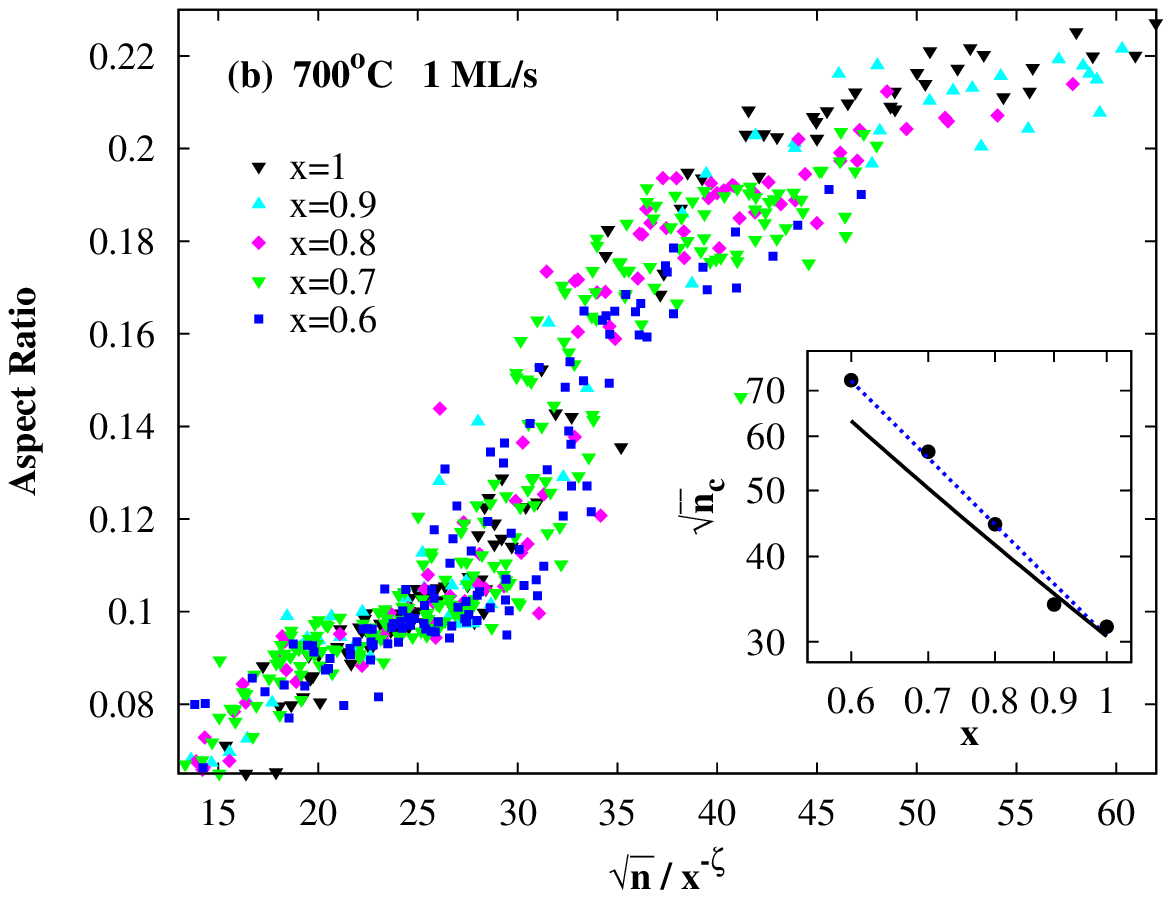}}
\caption{
    \label{F:aspect} 
    (a) Plot of island aspect ratio $r$ vs root of island size
    $\sqrt{n}$ for various $T$ and $R$ from simulations of deposition
    with lattice width $2048$. (b) Plot of $r$ vs
    $\sqrt{n}/x^{-\zeta}$ with $\zeta=1.69$ for various $x$ from
    similar simulations. Inset: Log-log plot of $\sqrt{n_c}$ vs
    $x$ where $n_c$ is the transition island size. The dotted line
    shows a linear fit to the data giving $\zeta=1.69$. The solid
    line represents a theoretical result.  }
\end{figure}

Large scale deposition simulations have been performed at
temperature $T$ from 450$^\circ$C to 850$^\circ$C at $x=1$ on lattices of width 2048.  The deposition rate
$R$ varies from 0.006 to 8 ML/s chosen empirically to generate
typically 3 to 5 pyramids or domes on each substrate.
The low island density minimizes elastic interactions among islands
which are known to alter the dynamics \cite{Floro2000,Capellini2003}.
An island is defined as one in which all constituent columns must be
at least 2 atoms tall.  We measure island size in number of atoms $n$
so that $\sqrt{n}$ is proportional to the linear size of the island.
Also, island aspect ratio is defined by $r= h / 2l$ where $h$ is the
height of the highest point of the island and $2l$ is its basewidth.
Figure \ref{F:aspect}(a) shows a scatter plot of the aspect ratio $r$
against $\sqrt{n}$ for all islands from 3 independent runs at each
temperature. Measurements are conducted throughout the evolution. Time
averaging of values associated with individual islands over
short periods are performed, but no ensemble averaging is done as each
island develops in general at a different pace.
We observe that $r$ first converges towards 0.1 as islands transform from stepped mounds into pyramids. It then rises again to around 0.2 characterizing the dome transition similar to experiment findings in Ref. \cite{Montalenti2004}. 
The much lower density of the data points at $0.1 < r < 0.2$ corresponding to highly unstable intermediate states was also observed in Ref. \cite{Montalenti2004}.
The morphologies of these intermediate states have been shown in Fig. 2(b). 
Results in Fig. 3 reveals two
distinct trends.  For the runs at $T \agt 650 ^ \circ C$, all islands
follows an identical evolution path with the dome transition occurring
at size ${n_c}\simeq 900$. In contrast, at lower temperature $T\alt
600 ^\circ C$, the transitions are delayed randomly to increasingly
larger sizes. At $T=450^\circ$ for instance, $n_c$ ranges from about
$900$ to $2000$. We will explain these distinct trends at
the end of this section.

Similar deposition simulations are also performed at $T=700^\circ$C
and $R=1$ ML/s for Ge concentration $x$ from 0.6 to 1 with 3
independent runs in each case. 
%
We find that the dome transitions occur at increasingly larger island
sizes as $x$ decreases.
The precise dependence is easily illustrated by a rescaled plot of
 $r$ against $\sqrt{n} /
x^{-\zeta}$ with $\zeta=1.69$ as shown in Fig.
\ref{F:aspect}(b). Data for various values of $x$ collapse
reasonably well into a single curve, implying
\begin{equation}
  r = f(n/x^{-\zeta})
\end{equation}
where $f$ is a rescaled function. 
To calculate $\zeta$ used above, we have measured the transition size
$n_c$ by averaging the sizes of all islands right at the transitions
with $ 0.12 \le r \le 0.16$. The resulting plot of $\sqrt{n_c}$
against $x$ in log-log scales is shown in the inset of
Fig. \ref{F:aspect}(b). A linear relation observed in the log-log plot
implies
\begin{equation}
\label{nc_x}
\sqrt{n_c} \sim x^{-\zeta}
\end{equation}
and a linear fit gives $\zeta=1.69$.
 This scaling relation will
be explained in Sec. IV.

\begin{figure}[htp]
 \centerline{\epsfxsize 0.99\columnwidth \epsfbox{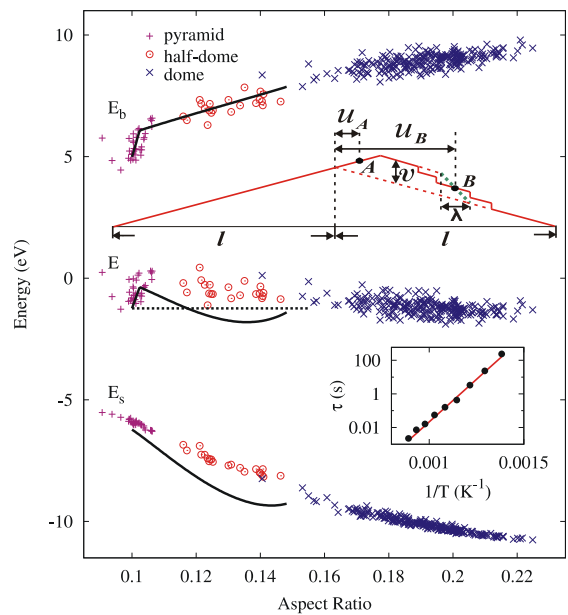}}
 \caption{
 \label{F:theory} 
 Plot of bond energy $E_b$, elastic energy $E_s$ and total energy $E$
 against island aspect ratio $r$ from the annealing of an initially
 pyramidal island on a substrate of width 256 at 550$^\circ$C
 (symbols). The solid lines show theoretical results. The schematic
 diagram shows a pyramid with additional layers on one of the shallow
 facets during the transition into a half-dome. 
 Inset: An Arrhenius plot of the dome
 transition time $\tau$.  }
\end{figure}

To study the energies of individual islands, we have performed simulations
on annealing of single pyramids directly constructed on smaller
substrates each of width $256$. The annealing temperature is 550$^\circ$C. 
The pyramid is initially
bounded by shallow facets and sits on an otherwise empty substrate
surface. It contains 1230 atoms and has a basewidth slightly less than
the lattice width. This number is chosen
empirically to be just sufficient to ensure an irreversible dome
transition.
 Figure \ref{F:theory} shows a
scatter plot of the system bond energy $E_b$, strain energy $E_s$ and
total energy $E=E_b+E_s$ against $r$ measured during the annealing
from 16 independent runs. 
Since a single island dominates, these energy of the whole system
approximates those of an island. 
Only time averaging of the values over short periods but
no ensemble averaging has been carried out. The symbols used indicate
if the data points correspond to pyramids, half-domes, or domes. The
geometries are identified reliably by the number of steep facets
present. The result indicates that there is an energy barrier
for the transition. Moreover, a number of data points
associated with half-domes cluster around $r \simeq 0.12-0.14$ showing
that the geometry characterizes a meta-stable state.

We next show that the dome transition is an activated process. 
We have repeated the above simulations on the annealing of pyramids at
$T$ from 450$^\circ$C to 850$^\circ$C. 
A dome transition time $\tau$ defined as the
average annealing duration required to reach an aspect ratio $r \ge
0.12$ is measured. 
Values of $\tau$ each averaged over 16 independent runs are
plotted against $1/T$ in the inset in Fig. \ref{F:theory}.  The data
fits well to 
\begin{equation}
\label{tau}
\tau = \tau_o \exp( \Omega_0/kT)
\end{equation}
 with $\Omega_0=1.97$~eV and $\tau_o
= 2.9 \times 10^{-12}$ s. The Arrhenius temperature dependence is
typical of activated processes. The
value of $\Omega_o$ will be explained in Sec. IV.
We further repeat the simulation 300 times at $700^\circ$C. The values
of $\tau$ measured are histogrammed. The result is well fitted by an
exponential distribution. This further supports that the dome
transition is an activated process.

With the shape transition time $\tau$ known, the distinct trends
followed by the data in Fig. \ref{F:aspect}(a) can now be explained. The
deposition rate $R$ has been empirically chosen to produce a constant
and sparse
island density. The choice hence ultimately depends on
pyramid nucleation and coarsening dynamics and is not directly related
to the
dome transition dynamics. For the $T \ge 650^\circ$C runs, we find
that $1/\tau \gg R$. The shape transition is thus fast compared with
deposition and hence also the island growth. The dome transition
process is only limited by the availability of atoms. The geometry as
characterized by the aspect ratio $r$ therefore only depends on
$\sqrt{n}$ and is independent of $T$ and $R$ as shown in
Fig. \ref{F:aspect}(a). In contrast, for the $T < 650^\circ$C runs,
we get $1/\tau \alt R$. Island growth can then be fast enough to
out-run the dome transition, which becomes rate-limited. There is a
significant random waiting time for the dome transition following an
exponential distribution during which the island may already have
grown to a larger size.  The transition thus occurs at a more broadly
distributed island size $n_c$ as observable from
Fig. \ref{F:aspect}(a). Note that if we consider for example a higher
island density by increasing the values of $R$ used, the
characteristic temperature separating the two trends, which is found
to be around $650^\circ C$ here, will increase.

\newcommand{\Eb}{E_b}

\section{Theory of shape transition}

We now present a detailed theoretical analysis based on generic forms
of elastic and facet energy for the transition of a pyramid into a
half-dome in 2D, which is applied to interpret our KMC simulation
results. Half-domes are meta-stable and they quickly transform into
domes. Our formulation is consistent with 
that of 
Montalenti {\it et al.} who have shown 
using energy parameters for Ge/Si from first-principle
calculations 
that the dome
transition is energetically favorable for sufficiently large pyramids
\cite{Montalenti2004}.

Consider an island of size $n$ initially in the form of a 
pyramid with a half-basewidth $l_0$. We have
\begin{equation}
  \label{n}
n=s_1l_0^2.  
\end{equation}
Geometrical rearrangements can lead to the formation new atomic layers of total vertical
thickness $v$ on one of the facets as shown in the 
schematic diagram in Fig. \ref{F:theory}.
The new half-basewidth $l$ is then related to $v$ by
\begin{equation}
  \label{n2}
n= s_1 l^2 + (u_B-u_A) v
\end{equation}
where $u_A$ and $u_B$ denote the positions of the midpoints A and B on
the edges of the new layers measured from the apex of the base
pyramid.  A single atomic step on a shallow facet has a height $s_1$.
We assume for simplicity that there are $v/s_1$ single steps at point
B.  Using Eq. (\ref{Eb}), the bond energy of the pyramid is
\begin{equation}
\label{py}
  E_b^{py}= 2 \phi_1 l + 2 \psi_{01} + \psi_{11}  + \beta_1 v/s_1.
\end{equation}
The pyramid becomes a half-dome when all the steps at point B turn
into a steep facet of $\lambda = v /
(s_2-s_1)$ columns wide. The bond energy of the resulting half-dome
also  follows from Eq. (\ref{Eb}) and we get 
\begin{equation}
\label{dome}
\Eb^{hdome} = 2 \phi_1 l + 2 \psi_{01} + \psi_{11}  +   \frac{
  \phi_2-\phi_1}{s_2-s_1} v + 2\psi_{12}
\end{equation}
Neglecting the small difference in the elastic energies of the two geometries,
the island takes the form with the lowest bond energy 
$E_b ={\mbox minimum}\{E_b^{py},E_b^{hdome}\}$.

The elastic energy of the pyramid and the half-dome is
assumed to be identical and is calculated by approximating both
edges of the new layers as vertical walls located at A and B.  A
shallow island approximation \cite{Tersoff1994,Tersoff1999} gives
\begin{equation}
  \label{eq:Es}
E_s = C \epsilon^2  \int\int dx dx' s(x)
s(x') \ln \frac{\mid x - x' \mid}{a_c},
\end{equation}
where  
$s(x)$ denotes the local surface slope of the island at position $x$
and $a_c$ is a
spatial cutoff. 
We put
$s(x) = - sgn(x) s_1 + v
\delta(x-u_A) - v \delta(x-u_B)$
%
where $sgn(x)$ and $\delta(x)$ represent the sign function and the
Dirac delta function.  In 2D, $C=\sigma_b^2a_s^2/\pi\epsilon^2 Y$ where
$\sigma_b \propto \epsilon$ is
the $xx$ component of the bulk misfit stress and $Y$ is the Young's
modulus. For our model, a simple calculation based on lattice elasticity gives $C = 4 k_n
a_s^2/3\pi$. 

Performing the integrations in Eq. (\ref{eq:Es}), we have
\begin{widetext}
\begin{equation}
  \label{Es}
  E_s = -2 C \epsilon^2\left\{ 2 \ln(2) s_1^2 l^2 +  s_1 v \sum_{p=A,B}\xi_P 
  \left[ l \ln \left(\frac{l+u_p}{l-u_p}\right) +
  u_p \ln \left(\frac{l^2}{u_p^2}-1 \right)\right] 
  + { v^2} \ln  \left(\frac{u_B-u_A}{a_c} \right)\right\}
\end{equation}
\end{widetext}
where $\xi_A=-1$ and $\xi_B=1$. In the following, we put $a_c =
e^{-3/2}\bar{\lambda}$ where $\bar{\lambda}=(\lambda+2u_A)/2$ is the
average spatial extent of the misfit force monopoles applied over the
edges at A and B. It can be shown that this choice gives the correct
energy when approaching the point force limit.

From simple geometry, $u_A=v/4s_1$. 
We calculate $u_B$ by 
minimizing the total
island energy $E$ at small $v$. Linearizing $E$ 
from Eqs. (\ref{n2})-(\ref{dome}) and (\ref{Es}) w.r.t. $v$, it can be shown
after some algebra that for both pyramids and
half-domes, $E$ is minimized at 
\begin{equation}
\label{uB}
{u}_B = l_0 \left [ 1 + 4 \exp\left (- \frac{\phi_1}{2 C
      \epsilon^2 s_1^2 l_0}\right) \right ]^{-1/2}.
\end{equation}
The total energy cost $\Delta E$ of an island compared with the
initial pyramid can then be
calculated. For a half-dome, we get, up to
linear order in $v$,
\begin{equation}
\label{DElinear}
\Delta E =  \left[ \frac{
   \phi_2-\phi_1}{s_2-s_1}
- 2 C \epsilon^2  s_1 l_0 \ln \left ( \frac{l_o+{u_B}}{l_o-{u_B}} \right )\right]
v 
+ 2 \psi_{12}.  
\end{equation}
Equation (\ref{uB}) gives the energetically most
favorable position for the initial formation of a steep facet.
Using the KMC model parameters for $x=1$, it gives $u_B/l_o
\simeq 0.53 $ for islands around the transition size. The
result is in general close to a limiting value $u_B/l_0=5 ^{-1/2}
\simeq 0.447$ obtained by neglecting the shallow facet formation energy $\phi_1$.


The energies $E_b$, $E_s$
and $E$ are numerically calculated for various layer thickness $v$
adopting the KMC model parameters for $x=1$
using Eqs. (\ref{n2})-(\ref{dome}) and (\ref{Es})-(\ref{uB}).
The island aspect ratio $r$ is also calculated as a function of $v$
using $r = s_1/2 + v/4l$ and Eq. (\ref{n2}).  In Fig. \ref{F:theory},
the energies are plotted as solid lines against $r$.  We have assumed
an island size ${n}= 1183$ atoms measured during the dome
transition in the KMC annealing simulations responsible for the data
points in Fig. \ref{F:theory}.  The only tunable parameter is a fitted
additive constant $3.4$~eV for $E_b$, which
accounts for the bond energies of all other excitations in the system.
It nevertheless plays no role in further calculations.
The theoretical estimates of the energies generally show reasonable
quantitative agreement with simulation results as observed in
Fig. \ref{F:theory}.  The main discrepancies are due to errors in $E_s$,
since the shallow island approximation is known to overestimate the
elastic relaxation at large $r$. Nevertheless, important features
including a shape transition energy barrier and a meta-stable
half-dome state are correctly reproduced and these will be further
studied.

From Fig. \ref{F:theory}, both theory and KMC simulation show an
energy barrier for the dome transition associated with a maximum in
the total energy $E$.  Its location follows theoretically from
$E^{py}=E^{hdome}$. We get a barrier height $\Delta
E_{max}=0.88$~eV which occurs at $r= 1.03$ or $v = 0.81$. It
corresponds to $v/s_1\simeq 4$ new atomic layers on the shallow
facet. For $r$ below and above 1.03 respectively, pyramid and
half-dome are the energetically preferred states.  Due to
the small value of $v$ at the barrier, $\Delta E_{max}$ is dominated
by the steep facet interface energy term as can be deduced from Eq. (\ref{DElinear}),
i.e. $\Delta E_{max} \simeq 2\psi_{12} = 1$~eV.  The
dominance of the steep facet interface energy on the transition energy
barrier may be a general feature applicable also in 3D.

%

The existence of an energy barrier confirms that the dome transition
is an activated process as suggested in Sec. III. The transition rate
$R$ hence follows the Arrhenius form $ R = \nu \exp(- \Delta
E_{max}/kT)$, where $\nu$ denotes an attempt frequency. Assuming that
the transition is limited by the diffusion of adatoms on the shallow
facet, one expect $\nu \propto \rho D$, where $\rho=\exp(-E_{ad}/kT)$
and $D \propto \exp(-\Omega_{ad}/kT)$ are the adatom density and
diffusion coefficient on the shallow facet.  Here, $E_{ad}=0.6$~eV and
$\Omega_{ad}=0.57$~eV are the adatom formation energy and hopping
energy barrier on the shallow facet calculated using Eqs. (\ref{Eb})
and (\ref{rate2}). In particular, $\Omega_{ad}$ is not far from a previous estimate from  first-principles calculations \cite{Montalenti2004b}. The dome transition time $\tau \propto 1/R$ is
hence given by
\begin{equation}
  \label{tau2}
\tau
\propto \exp\left(\frac{\Delta E_{max} + E_{ad} + \Omega_{ad}}{kT}\right)
\end{equation}
A comparison with Eq. (\ref{tau}) leads to
$\Omega_0 = \Delta E_{max} + E_{ad} + \Omega_{ad}$.  It gives
 $\Omega_o= 2.05$~eV in agreement with
$\Omega_0=1.97$~eV obtained above from KMC simulations.

\section{scaling of shape transition size}

We first assume quasi-equilibrium conditions in which the dome
transition is fast compared with island growth. It applies to the
cases of slow deposition and small transition barrier and is valid for
our KMC simulations at $T \agt 650^\circ C$ (see
Fig. \ref{F:aspect}(a)). The island energy
$E$ from theory as shown in Fig. \ref{F:theory} exhibits a local minimum representing
the meta-stable half-dome state. The energy rises again at larger $r$
because the base pyramid then becomes too small to relief the elastic
energy efficiently.  The dome transition occurs only if the
half-dome is sufficiently stable, say of energy $kT$ below that of the
initial pyramid. For island at the transition size $n_c$, the minimum
of $\Delta E$ hence follows $\Delta E_{min}=-kT$. We can then calculate
$n_c$ numerically using Eqs. (\ref{n2}),
(\ref{dome}), (\ref{Es}) and (\ref{uB}) and the values are plotted against $x$ as
a solid line in the inset of Fig. \ref{F:aspect}(b). Note that no
tunable parameter is involved. The values are in reasonable agreement
with the KMC results and supports the scaling relation in
Eq. (\ref{nc_x}) with $\zeta \simeq 1.49$ consistent with $1.69$
found in simulations.

\newcommand{\gtwo}{g_2}
\newcommand{\gthree}{g_3}

In addition to the numerical estimate of the exponent above, better
insights are obtained by deriving an exact exponent $\zeta=2$ valid
asymptotically in the small misfit limit, i.e. $\epsilon \propto x
\rightarrow 0$. Assume that the relative position of the steep facet
is independent of $\epsilon$ so that $u_B \propto l_o$, which will be
justified later. Simple scaling properties in 2D elasticity imply that
the change in the elastic energy of a half-dome compared with the initial
pyramid follows $\Delta E_s = - \epsilon^2 l_0^2 \gtwo(v/l_o)$ for
some function $\gtwo$. This is also explicitly derivable from
Eq. (\ref{Es}).
The total
energy cost of a half-dome is hence
\begin{equation}
\label{DE2}
\Delta E = A_2  v - \epsilon^2 l_0 ^2 \gtwo(v/l_o) + B_2
\end{equation}
where $A_2 = ({\phi_2-\phi_1})/({s_2-s_1})$ and $B_2=2 \psi_{12}$.  At
island transition size $n_c$ and considering a layer height $v$
minimizing $\Delta E$ to $\Delta E_{min}=-kT$, we have
\begin{equation}
\label{DE3}
\epsilon^2 l_o^2 \gtwo(v/l_o) - A_2 v =  B_2 + kT
\end{equation}
It means that the elastic energy gain must overcompensate the facet
formation energy cost by an excess amount $B_2 + kT$. As $\epsilon
\rightarrow 0$, we will see in the following that the shape transition occurs at a
larger island size. Both energy terms on the l.h.s. of Eq. (\ref{DE3})
increase unboundedly and must balance each other, while the constant
energy excess becomes negligible.  Therefore, we have $\epsilon^2
l_o^2 \gtwo(v/l_o) = A v$. The meta-stable half-dome state at 
transition is thus characterized by the scaling solution $l_o \sim v \sim
\epsilon^{-2}$. Using $\sqrt{n_c} \propto l_0$ and $\epsilon=0.04 x$,
we get
\begin{equation}
\label{scaling}
\sqrt{n_c} \sim x^{-2}
\end{equation}
i.e. ${\zeta=2}$. This solution is consistent with the assumption
that $u_B$ is independent of $\epsilon$ as deduced using
Eq. (\ref{uB}). It also justifies that both terms on the l.h.s. of Eq. (\ref{DE3})
increase unboundedly as $\epsilon \rightarrow 0$.

For finite $\epsilon$ and $x$, the energy excess $B_2 + kT$ in
Eq. (\ref{DE3}) is not negligible. It gives a finite size correction
to the exact scaling in Eq. (\ref{scaling}). This results at an
effective scaling $\sqrt{n_c} \sim x^{-\zeta}$ with $\zeta \alt 2$
consistent with $\zeta =1.69$ found in our KMC simulation.

In 3D, the dome transition is initiated by step bunching close to the
mid-level on a pyramid \cite{Montalenti2004}. Generalizing our
discussion above, we consider the formation of a square ring of steep
facet of vertical thickness $v$ on a 3D pyramid of basewidth $2l_o$. Generalizing 
Eq. (\ref{DE2}), the energy cost is
\begin{equation}
\label{DE3d}
\Delta E = A_3  v l_o - \epsilon^2 l_0 ^3 \gthree(v/l_o) + B_3 l_o
\end{equation}
where the terms on the r.h.s. similarly denote the facet formation energy,
the elastic energy gain, and the steep facet interface energy respectively,
for some smooth function $g_3$ and constants $A_3$ and $B_3$
independent of $v$ and $l_o$. 
A similar calculation leads to $n_c^{1/3}
\propto l_o \sim x^{-2}$. 
This generalizes Eq.
(\ref{scaling}) to 
\begin{equation}
\label{scaling2}
{n_c}^{1/d} \sim  x^{-\zeta}
\end{equation}
with the same exponent $\zeta=2$ in dimension $d=2$ or 3 for $x\rightarrow 0$.
For finite $x$, an effective exponent $\zeta\alt 2$ is
expected in both dimensions.

The asymptotic scalings derived above for quasi-equilibrium conditions
essentially follows from the balance between the steep facet formation
energy and the elastic energy gain which scales with the island size
differently. It is analogous to the scaling predicted for island
formation size based on the Asaro-Tiller-Grinfeld instability theory
\cite{Politi2000}. Due to the simplicity, it is also rather robust as
will be shown below.

Instead of assuming quasi-equilibrium conditions, the dome transition
can be limited by the kinetics. This may be appropriate in particular
at 3D for small $x$ since the barrier predicted above can become too
large to overcome. The transition is then delayed to a larger island
size which lowers the barrier.  Let us then assume a very simple
transition criterion that the energy barrier $\Delta E_{max}$ must not
exceed a given value, say a few times of $kT$. A similar calculation
for $x \rightarrow 0$ again arrives at Eq. (\ref{scaling2}) with the
same exponent $\zeta=2$ in both 2D and 3D. The solution also requires
$l_o \sim \epsilon^{-2}$ but  $v \sim
\epsilon^0$ and we have used $g_d(z) \propto z$ for $z\rightarrow 0$
which readily follows in the 2D case from Eq. (\ref{DElinear}) after neglecting
a logarithmic factor.


\section{discussions}

In the calculations above, $x$ denotes the actual Ge concentration in
the film so that $\epsilon=0.04 x$. In experiments, it can differ
greatly from the nominal concentration due to intermixing with
substrate atoms and this complicates interpretation of experiment
results. As a very rough estimate, experiments on deposition of pure
Ge at e.g. 450$^\circ$C and 700$^\circ$C have found dome transition
occurring at island volume $2800$ nm$^3$ \cite{Drucker2002} and $2
\times 10^5$ nm$^3$ \cite{Rastelli2006}. Neglecting compositional
non-uniformity, it has been estimated that the actual Ge concentrate
in the islands is $x = 0.82$ and 0.43 respectively at 450$^\circ$C and
700$^\circ$C \cite{Drucker2003}. This gives a very preliminary
estimate for the scaling exponent $\zeta\simeq 2.2$ consistent with
the asymptotic value $\zeta= 2$ derived above, although more
experiments are required for a reliable conclusion.

Our simulations and theoretical calculations show that occurrence of
well-defined dome transitions depends strongly in particular on the
formation and interface energies of steep facets. The detailed
dependences of these energies on the Ge concentration and temperature
are neglected. We have also neglected the spatial non-uniformity of Ge
concentration, surface stress and realistic crystal elastic
anisotropy. They should have significant quantitative impacts on the
shape transition, but are not expected to alter the dynamics
described here qualitatively.

In summary, we have generalized a multi-state lattice model for
elastic solids to account for both shallow and steep facets with
tunable energy parameters. Using this model, we perform kinetic Monte
Carlo simulations to study the pyramid-to-dome transition in the
heteroepitaxy growth of Ge$_{x}$Si$_{1-x}$ on Si in 2D.  For sufficiently slow
deposition, the shape transition occurs at an island size independent
of temperature and deposition rate. A scaling relation between the
transition size and the Ge concentration is observed. For fast
deposition, the transition can be delayed randomly to a larger island
size. For annealing simulations, the shape transition time is found to
follow an Arrhenius form. A theory based on elastic energy in the
shallow island approximation and simple forms of facet formation
energies is derived. Numerical solutions of the energetic equations
give island energies, shape transition size, and shape transition rate
in reasonable agreement with simulations. The shape transition energy
barrier is dominated by the interface energy between the shallow and
the steep facets.  We have also derived analytically an exact scaling
rule between the transition size and the Ge concentration applicable
in the limit of small Ge concentration which is expected to be valid
in both 2D and 3D. A finite size correction to the scaling at higher
Ge concentration is explained.

This work was supported by HK GRF, Grant No. PolyU-5009/06P and PolyU
Grant No. G-U354. 

\bibliography{island}

\begin{thebibliography}{37}
\expandafter\ifx\csname natexlab\endcsname\relax\def\natexlab#1{#1}\fi
\expandafter\ifx\csname bibnamefont\endcsname\relax
  \def\bibnamefont#1{#1}\fi
\expandafter\ifx\csname bibfnamefont\endcsname\relax
  \def\bibfnamefont#1{#1}\fi
\expandafter\ifx\csname citenamefont\endcsname\relax
  \def\citenamefont#1{#1}\fi
\expandafter\ifx\csname url\endcsname\relax
  \def\url#1{\texttt{#1}}\fi
\expandafter\ifx\csname urlprefix\endcsname\relax\def\urlprefix{URL }\fi
\providecommand{\bibinfo}[2]{#2}
\providecommand{\eprint}[2][]{\url{#2}}

\bibitem[{\citenamefont{Politi et~al.}(2000)\citenamefont{Politi, Grenet,
  Marty, Ponchet, and Villain}}]{Politi2000}
\bibinfo{author}{\bibfnamefont{P.}~\bibnamefont{Politi}},
  \bibinfo{author}{\bibfnamefont{G.}~\bibnamefont{Grenet}},
  \bibinfo{author}{\bibfnamefont{A.}~\bibnamefont{Marty}},
  \bibinfo{author}{\bibfnamefont{A.}~\bibnamefont{Ponchet}}, \bibnamefont{and}
  \bibinfo{author}{\bibfnamefont{J.}~\bibnamefont{Villain}},
  \bibinfo{journal}{Phys. Rep.} \textbf{\bibinfo{volume}{324}},
  \bibinfo{pages}{271} (\bibinfo{year}{2000}).

\bibitem[{\citenamefont{Shchukin et~al.}(2003)\citenamefont{Shchukin,
  Ledentsov, and Bimberg}}]{Shchukin2003}
\bibinfo{author}{\bibfnamefont{V.~A.} \bibnamefont{Shchukin}},
  \bibinfo{author}{\bibfnamefont{N.~N.} \bibnamefont{Ledentsov}},
  \bibnamefont{and} \bibinfo{author}{\bibfnamefont{D.}~\bibnamefont{Bimberg}},
  \emph{\bibinfo{title}{Epitaxy of nanostructures}}
  (\bibinfo{publisher}{Springer}, \bibinfo{year}{2003}).

\bibitem[{\citenamefont{Berbezier and Ronda}(2009)}]{Berbezier2009}
\bibinfo{author}{\bibfnamefont{I.}~\bibnamefont{Berbezier}} \bibnamefont{and}
  \bibinfo{author}{\bibfnamefont{A.}~\bibnamefont{Ronda}},
  \bibinfo{journal}{Surf. Sci. Rep.} \textbf{\bibinfo{volume}{64}},
  \bibinfo{pages}{47} (\bibinfo{year}{2009}).

\bibitem[{\citenamefont{Ross et~al.}(1999)\citenamefont{Ross, Tromp, and
  Reuter}}]{Ross1999}
\bibinfo{author}{\bibfnamefont{F.~M.} \bibnamefont{Ross}},
  \bibinfo{author}{\bibfnamefont{R.~M.} \bibnamefont{Tromp}}, \bibnamefont{and}
  \bibinfo{author}{\bibfnamefont{M.~C.} \bibnamefont{Reuter}},
  \bibinfo{journal}{Science} \textbf{\bibinfo{volume}{286}},
  \bibinfo{pages}{1931} (\bibinfo{year}{1999}).

\bibitem[{\citenamefont{Ross et~al.}(1998)\citenamefont{Ross, Tersoff, and
  Tromp}}]{Ross1998}
\bibinfo{author}{\bibfnamefont{F.~M.} \bibnamefont{Ross}},
  \bibinfo{author}{\bibfnamefont{J.}~\bibnamefont{Tersoff}}, \bibnamefont{and}
  \bibinfo{author}{\bibfnamefont{R.~M.} \bibnamefont{Tromp}},
  \bibinfo{journal}{Phys. Rev. Lett.} \textbf{\bibinfo{volume}{80}},
  \bibinfo{pages}{984} (\bibinfo{year}{1998}).

\bibitem[{\citenamefont{Montalenti
  et~al.}(2004{\natexlab{a}})\citenamefont{Montalenti, Raiteri, Migas, von
  Kanel, Rastelli, Manzano, Costantini, Denker, Schmidt, Kern
  et~al.}}]{Montalenti2004}
\bibinfo{author}{\bibfnamefont{F.}~\bibnamefont{Montalenti}},
  \bibinfo{author}{\bibfnamefont{P.}~\bibnamefont{Raiteri}},
  \bibinfo{author}{\bibfnamefont{D.}~\bibnamefont{Migas}},
  \bibinfo{author}{\bibfnamefont{H.}~\bibnamefont{von Kanel}},
  \bibinfo{author}{\bibfnamefont{A.}~\bibnamefont{Rastelli}},
  \bibinfo{author}{\bibfnamefont{C.}~\bibnamefont{Manzano}},
  \bibinfo{author}{\bibfnamefont{G.}~\bibnamefont{Costantini}},
  \bibinfo{author}{\bibfnamefont{U.}~\bibnamefont{Denker}},
  \bibinfo{author}{\bibfnamefont{O.}~\bibnamefont{Schmidt}},
  \bibinfo{author}{\bibfnamefont{K.}~\bibnamefont{Kern}}, \bibnamefont{et~al.},
  \bibinfo{journal}{Phys. Rev. Lett.} \textbf{\bibinfo{volume}{93}},
  \bibinfo{pages}{216102} (\bibinfo{year}{2004}{\natexlab{a}}).

\bibitem[{\citenamefont{Cereda and Montalenti}(2007)}]{Montalenti2007}
\bibinfo{author}{\bibfnamefont{S.}~\bibnamefont{Cereda}} \bibnamefont{and}
  \bibinfo{author}{\bibfnamefont{F.}~\bibnamefont{Montalenti}},
  \bibinfo{journal}{Phys. Rev. B} \textbf{\bibinfo{volume}{75}},
  \bibinfo{pages}{195321} (\bibinfo{year}{2007}).

\bibitem[{\citenamefont{Fujikawa et~al.}(2002)\citenamefont{Fujikawa, Akiyama,
  Nagao, Sakurai, Lagally, Hashimoto, Morikawa, and Terakura}}]{Fujikawa2002}
\bibinfo{author}{\bibfnamefont{Y.}~\bibnamefont{Fujikawa}},
  \bibinfo{author}{\bibfnamefont{K.}~\bibnamefont{Akiyama}},
  \bibinfo{author}{\bibfnamefont{T.}~\bibnamefont{Nagao}},
  \bibinfo{author}{\bibfnamefont{T.}~\bibnamefont{Sakurai}},
  \bibinfo{author}{\bibfnamefont{M.~G.} \bibnamefont{Lagally}},
  \bibinfo{author}{\bibfnamefont{T.}~\bibnamefont{Hashimoto}},
  \bibinfo{author}{\bibfnamefont{Y.}~\bibnamefont{Morikawa}}, \bibnamefont{and}
  \bibinfo{author}{\bibfnamefont{K.}~\bibnamefont{Terakura}},
  \bibinfo{journal}{Phys. Rev. Lett.} \textbf{\bibinfo{volume}{88}},
  \bibinfo{pages}{176101} (\bibinfo{year}{2002}).

\bibitem[{\citenamefont{Cereda et~al.}(2005)\citenamefont{Cereda, Montalenti,
  and Miglio}}]{Montalenti2005}
\bibinfo{author}{\bibfnamefont{S.}~\bibnamefont{Cereda}},
  \bibinfo{author}{\bibfnamefont{F.}~\bibnamefont{Montalenti}},
  \bibnamefont{and} \bibinfo{author}{\bibfnamefont{L.}~\bibnamefont{Miglio}},
  \bibinfo{journal}{Surf. Sci.} \textbf{\bibinfo{volume}{591}},
  \bibinfo{pages}{23} (\bibinfo{year}{2005}).

\bibitem[{\citenamefont{Lu et~al.}(2005)\citenamefont{Lu, Cuma, and
  Liu}}]{Lu2005}
\bibinfo{author}{\bibfnamefont{G.-H.} \bibnamefont{Lu}},
  \bibinfo{author}{\bibfnamefont{M.}~\bibnamefont{Cuma}}, \bibnamefont{and}
  \bibinfo{author}{\bibfnamefont{F.}~\bibnamefont{Liu}},
  \bibinfo{journal}{Phys. Rev. B} \textbf{\bibinfo{volume}{72}},
  \bibinfo{pages}{125415} (\bibinfo{year}{2005}).

\bibitem[{\citenamefont{Shklyaev et~al.}(2005)\citenamefont{Shklyaev, Beck,
  Asta, Miksis, and Voorhees}}]{Shklyaev2005}
\bibinfo{author}{\bibfnamefont{O.}~\bibnamefont{Shklyaev}},
  \bibinfo{author}{\bibfnamefont{M.}~\bibnamefont{Beck}},
  \bibinfo{author}{\bibfnamefont{M.}~\bibnamefont{Asta}},
  \bibinfo{author}{\bibfnamefont{M.}~\bibnamefont{Miksis}}, \bibnamefont{and}
  \bibinfo{author}{\bibfnamefont{P.}~\bibnamefont{Voorhees}},
  \bibinfo{journal}{Phys. Rev. Lett.} \textbf{\bibinfo{volume}{94}},
  \bibinfo{pages}{176102} (\bibinfo{year}{2005}).

\bibitem[{\citenamefont{Makeev et~al.}(2003)\citenamefont{Makeev, Yu, and
  Madhukar}}]{Maxim2003}
\bibinfo{author}{\bibfnamefont{M.~A.} \bibnamefont{Makeev}},
  \bibinfo{author}{\bibfnamefont{W.}~\bibnamefont{Yu}}, \bibnamefont{and}
  \bibinfo{author}{\bibfnamefont{A.}~\bibnamefont{Madhukar}},
  \bibinfo{journal}{Phys. Rev. B} \textbf{\bibinfo{volume}{68}},
  \bibinfo{pages}{195301} (\bibinfo{year}{2003}).

\bibitem[{\citenamefont{Retford et~al.}(2007)\citenamefont{Retford, Asta,
  Miksis, Voorhees, and {Webb III}}}]{Retford2007}
\bibinfo{author}{\bibfnamefont{C.~M.} \bibnamefont{Retford}},
  \bibinfo{author}{\bibfnamefont{M.}~\bibnamefont{Asta}},
  \bibinfo{author}{\bibfnamefont{M.~J.} \bibnamefont{Miksis}},
  \bibinfo{author}{\bibfnamefont{P.~W.} \bibnamefont{Voorhees}},
  \bibnamefont{and} \bibinfo{author}{\bibfnamefont{E.~B.} \bibnamefont{{Webb
  III}}}, \bibinfo{journal}{Phys. Rev. B} \textbf{\bibinfo{volume}{75}},
  \bibinfo{pages}{075311} (\bibinfo{year}{2007}).

\bibitem[{\citenamefont{Tu and Tersoff}(2004)}]{Tu2004}
\bibinfo{author}{\bibfnamefont{Y.}~\bibnamefont{Tu}} \bibnamefont{and}
  \bibinfo{author}{\bibfnamefont{J.}~\bibnamefont{Tersoff}},
  \bibinfo{journal}{Phys. Rev. Lett.} \textbf{\bibinfo{volume}{93}},
  \bibinfo{pages}{216101} (\bibinfo{year}{2004}).

\bibitem[{\citenamefont{Huang et~al.}(2007)\citenamefont{Huang, Zhou, and
  h.~Chiu}}]{Chiu2007}
\bibinfo{author}{\bibfnamefont{Z.}~\bibnamefont{Huang}},
  \bibinfo{author}{\bibfnamefont{T.}~\bibnamefont{Zhou}}, \bibnamefont{and}
  \bibinfo{author}{\bibfnamefont{C.}~\bibnamefont{h.~Chiu}},
  \bibinfo{journal}{Phys. Rev. Lett.} \textbf{\bibinfo{volume}{98}},
  \bibinfo{eid}{196102} (\bibinfo{year}{2007}).

\bibitem[{\citenamefont{Orr et~al.}(1992)\citenamefont{Orr, Kessler, Snyder,
  and Sander}}]{Orr1992}
\bibinfo{author}{\bibfnamefont{B.~G.} \bibnamefont{Orr}},
  \bibinfo{author}{\bibfnamefont{D.}~\bibnamefont{Kessler}},
  \bibinfo{author}{\bibfnamefont{C.~W.} \bibnamefont{Snyder}},
  \bibnamefont{and} \bibinfo{author}{\bibfnamefont{L.}~\bibnamefont{Sander}},
  \bibinfo{journal}{Europhys. Lett.} \textbf{\bibinfo{volume}{19}},
  \bibinfo{pages}{33} (\bibinfo{year}{1992}).

\bibitem[{\citenamefont{Lam et~al.}(2002)\citenamefont{Lam, Lee, and
  Sander}}]{Lam2002}
\bibinfo{author}{\bibfnamefont{C.-H.} \bibnamefont{Lam}},
  \bibinfo{author}{\bibfnamefont{C.-K.} \bibnamefont{Lee}}, \bibnamefont{and}
  \bibinfo{author}{\bibfnamefont{L.~M.} \bibnamefont{Sander}},
  \bibinfo{journal}{Phys. Rev. Lett.} \textbf{\bibinfo{volume}{89}},
  \bibinfo{pages}{216102} (\bibinfo{year}{2002}).

\bibitem[{\citenamefont{Gray et~al.}(2005)\citenamefont{Gray, Hull, Lam,
  Sutter, Means, and Floro}}]{Gray2005}
\bibinfo{author}{\bibfnamefont{J.~L.} \bibnamefont{Gray}},
  \bibinfo{author}{\bibfnamefont{R.}~\bibnamefont{Hull}},
  \bibinfo{author}{\bibfnamefont{C.-H.} \bibnamefont{Lam}},
  \bibinfo{author}{\bibfnamefont{P.}~\bibnamefont{Sutter}},
  \bibinfo{author}{\bibfnamefont{J.}~\bibnamefont{Means}}, \bibnamefont{and}
  \bibinfo{author}{\bibfnamefont{J.~A.} \bibnamefont{Floro}},
  \bibinfo{journal}{Phys. Rev. B} \textbf{\bibinfo{volume}{72}},
  \bibinfo{pages}{155323} (\bibinfo{year}{2005}).

\bibitem[{\citenamefont{Schulze and Smereka}(2009)}]{Schulze2009}
\bibinfo{author}{\bibfnamefont{T.~P.} \bibnamefont{Schulze}} \bibnamefont{and}
  \bibinfo{author}{\bibfnamefont{P.}~\bibnamefont{Smereka}},
  \bibinfo{journal}{J. Mech. Phys. Solids} \textbf{\bibinfo{volume}{57}},
  \bibinfo{pages}{521 } (\bibinfo{year}{2009}).

\bibitem[{\citenamefont{Lung et~al.}(2005)\citenamefont{Lung, Lam, and
  Sander}}]{Lung2005}
\bibinfo{author}{\bibfnamefont{M.~T.} \bibnamefont{Lung}},
  \bibinfo{author}{\bibfnamefont{C.-H.} \bibnamefont{Lam}}, \bibnamefont{and}
  \bibinfo{author}{\bibfnamefont{L.~M.} \bibnamefont{Sander}},
  \bibinfo{journal}{Phys. Rev. Lett.} \textbf{\bibinfo{volume}{95}},
  \bibinfo{pages}{086102} (\bibinfo{year}{2005}).

\bibitem[{\citenamefont{Russo and Smereka}(2006)}]{Smereka2006}
\bibinfo{author}{\bibfnamefont{G.}~\bibnamefont{Russo}} \bibnamefont{and}
  \bibinfo{author}{\bibfnamefont{P.}~\bibnamefont{Smereka}},
  \bibinfo{journal}{J. Comput. Phys.} \textbf{\bibinfo{volume}{214}},
  \bibinfo{pages}{809 } (\bibinfo{year}{2006}).

\bibitem[{\citenamefont{Lam and Lung}(2007)}]{Lam2007}
\bibinfo{author}{\bibfnamefont{C.-H.} \bibnamefont{Lam}} \bibnamefont{and}
  \bibinfo{author}{\bibfnamefont{M.~T.} \bibnamefont{Lung}},
  \bibinfo{journal}{Int. J. Mod. Phys. B} \textbf{\bibinfo{volume}{21}},
  \bibinfo{pages}{4219} (\bibinfo{year}{2007}).

\bibitem[{\citenamefont{Lam et~al.}(2008)\citenamefont{Lam, Lung, and
  Sander}}]{Lam2008}
\bibinfo{author}{\bibfnamefont{C.-H.} \bibnamefont{Lam}},
  \bibinfo{author}{\bibfnamefont{M.~T.} \bibnamefont{Lung}}, \bibnamefont{and}
  \bibinfo{author}{\bibfnamefont{L.~M.} \bibnamefont{Sander}},
  \bibinfo{journal}{J. Sci. Comput.} \textbf{\bibinfo{volume}{37}},
  \bibinfo{pages}{73} (\bibinfo{year}{2008}).

\bibitem[{\citenamefont{Lee et~al.}(2009)\citenamefont{Lee, Noordhoek, Smereka,
  McKay, and Millunchick}}]{Smereka2009}
\bibinfo{author}{\bibfnamefont{J.~Y.} \bibnamefont{Lee}},
  \bibinfo{author}{\bibfnamefont{M.~J.} \bibnamefont{Noordhoek}},
  \bibinfo{author}{\bibfnamefont{P.}~\bibnamefont{Smereka}},
  \bibinfo{author}{\bibfnamefont{H.}~\bibnamefont{McKay}}, \bibnamefont{and}
  \bibinfo{author}{\bibfnamefont{J.~M.} \bibnamefont{Millunchick}},
  \bibinfo{journal}{Nanotechnology} \textbf{\bibinfo{volume}{20}},
  \bibinfo{pages}{285305} (\bibinfo{year}{2009}).

\bibitem[{\citenamefont{Lam}(2010)}]{Lam2010}
\bibinfo{author}{\bibfnamefont{C.-H.} \bibnamefont{Lam}},
  \bibinfo{journal}{Phys. Rev. E} \textbf{\bibinfo{volume}{81}},
  \bibinfo{pages}{021607} (\bibinfo{year}{2010}).

\bibitem[{\citenamefont{Baskaran et~al.}(2010)\citenamefont{Baskaran, Devita,
  and Smereka}}]{Smereka2010}
\bibinfo{author}{\bibfnamefont{A.}~\bibnamefont{Baskaran}},
  \bibinfo{author}{\bibfnamefont{J.}~\bibnamefont{Devita}}, \bibnamefont{and}
  \bibinfo{author}{\bibfnamefont{P.}~\bibnamefont{Smereka}},
  \bibinfo{journal}{Cont. Mech. Thermo.} \textbf{\bibinfo{volume}{22}},
  \bibinfo{pages}{1} (\bibinfo{year}{2010}).

\bibitem[{\citenamefont{Ratsch et~al.}(1996)\citenamefont{Ratsch, Smilauer,
  Vvedensky, and Zangwill}}]{Ratsch1996}
\bibinfo{author}{\bibfnamefont{C.}~\bibnamefont{Ratsch}},
  \bibinfo{author}{\bibfnamefont{P.}~\bibnamefont{Smilauer}},
  \bibinfo{author}{\bibfnamefont{D.}~\bibnamefont{Vvedensky}},
  \bibnamefont{and} \bibinfo{author}{\bibfnamefont{A.}~\bibnamefont{Zangwill}},
  \bibinfo{journal}{J. Phys. I} \textbf{\bibinfo{volume}{6}},
  \bibinfo{pages}{575} (\bibinfo{year}{1996}).

\bibitem[{\citenamefont{Meixner et~al.}({2002})\citenamefont{Meixner, Scholl,
  Shchukin, and Bimberg}}]{Meixner2001}
\bibinfo{author}{\bibfnamefont{M.}~\bibnamefont{Meixner}},
  \bibinfo{author}{\bibfnamefont{E.}~\bibnamefont{Scholl}},
  \bibinfo{author}{\bibfnamefont{V.}~\bibnamefont{Shchukin}}, \bibnamefont{and}
  \bibinfo{author}{\bibfnamefont{D.}~\bibnamefont{Bimberg}},
  \bibinfo{journal}{{Phys. Rev. Lett.}} \textbf{\bibinfo{volume}{{87}}},
  \bibinfo{pages}{{236101}} (\bibinfo{year}{{2002}}).

\bibitem[{\citenamefont{Zhu et~al.}(2007)\citenamefont{Zhu, Pan, and
  Chung}}]{Zhu2007}
\bibinfo{author}{\bibfnamefont{R.}~\bibnamefont{Zhu}},
  \bibinfo{author}{\bibfnamefont{E.}~\bibnamefont{Pan}}, \bibnamefont{and}
  \bibinfo{author}{\bibfnamefont{P.~W.} \bibnamefont{Chung}},
  \bibinfo{journal}{Phys. Rev. B} \textbf{\bibinfo{volume}{75}},
  \bibinfo{pages}{205339} (\bibinfo{year}{2007}).

\bibitem[{\citenamefont{Floro et~al.}(2000)\citenamefont{Floro, Sinclair,
  Chason, Freund, Twesten, Hwang, and Lucadamo}}]{Floro2000}
\bibinfo{author}{\bibfnamefont{J.~A.} \bibnamefont{Floro}},
  \bibinfo{author}{\bibfnamefont{M.~B.} \bibnamefont{Sinclair}},
  \bibinfo{author}{\bibfnamefont{E.}~\bibnamefont{Chason}},
  \bibinfo{author}{\bibfnamefont{L.~B.} \bibnamefont{Freund}},
  \bibinfo{author}{\bibfnamefont{R.~D.} \bibnamefont{Twesten}},
  \bibinfo{author}{\bibfnamefont{R.~Q.} \bibnamefont{Hwang}}, \bibnamefont{and}
  \bibinfo{author}{\bibfnamefont{G.~A.} \bibnamefont{Lucadamo}},
  \bibinfo{journal}{Phys. Rev. Lett.} \textbf{\bibinfo{volume}{84}},
  \bibinfo{pages}{701} (\bibinfo{year}{2000}).

\bibitem[{\citenamefont{Capellini et~al.}(2003)\citenamefont{Capellini, Seta,
  and Evangelisti}}]{Capellini2003}
\bibinfo{author}{\bibfnamefont{G.}~\bibnamefont{Capellini}},
  \bibinfo{author}{\bibfnamefont{M.~D.} \bibnamefont{Seta}}, \bibnamefont{and}
  \bibinfo{author}{\bibfnamefont{F.}~\bibnamefont{Evangelisti}},
  \bibinfo{journal}{J. Appl. Phys.} \textbf{\bibinfo{volume}{93}},
  \bibinfo{pages}{291} (\bibinfo{year}{2003}).

\bibitem[{\citenamefont{Tersoff and LeGoues}(1994)}]{Tersoff1994}
\bibinfo{author}{\bibfnamefont{J.}~\bibnamefont{Tersoff}} \bibnamefont{and}
  \bibinfo{author}{\bibfnamefont{F.~K.} \bibnamefont{LeGoues}},
  \bibinfo{journal}{Phys. Rev. Lett.} \textbf{\bibinfo{volume}{72}},
  \bibinfo{pages}{3570} (\bibinfo{year}{1994}).

\bibitem[{\citenamefont{Daruka et~al.}(1999)\citenamefont{Daruka, Tersoff, and
  Barab\'asi}}]{Tersoff1999}
\bibinfo{author}{\bibfnamefont{I.}~\bibnamefont{Daruka}},
  \bibinfo{author}{\bibfnamefont{J.}~\bibnamefont{Tersoff}}, \bibnamefont{and}
  \bibinfo{author}{\bibfnamefont{A.-L.} \bibnamefont{Barab\'asi}},
  \bibinfo{journal}{Phys. Rev. Lett.} \textbf{\bibinfo{volume}{82}},
  \bibinfo{pages}{2753} (\bibinfo{year}{1999}).

\bibitem[{\citenamefont{Montalenti
  et~al.}(2004{\natexlab{b}})\citenamefont{Montalenti, Migas, Gamba, and
  Miglio}}]{Montalenti2004b}
\bibinfo{author}{\bibfnamefont{F.}~\bibnamefont{Montalenti}},
  \bibinfo{author}{\bibfnamefont{D.~B.} \bibnamefont{Migas}},
  \bibinfo{author}{\bibfnamefont{F.}~\bibnamefont{Gamba}}, \bibnamefont{and}
  \bibinfo{author}{\bibfnamefont{L.}~\bibnamefont{Miglio}},
  \bibinfo{journal}{Phys. Rev. B} \textbf{\bibinfo{volume}{70}},
  \bibinfo{pages}{245315} (\bibinfo{year}{2004}{\natexlab{b}}).

\bibitem[{\citenamefont{Drucker}(2002)}]{Drucker2002}
\bibinfo{author}{\bibfnamefont{J.}~\bibnamefont{Drucker}}, \bibinfo{journal}{J.
  Quant. Elect.} \textbf{\bibinfo{volume}{38}}, \bibinfo{pages}{975}
  (\bibinfo{year}{2002}).

\bibitem[{\citenamefont{Stoffel et~al.}(2006)\citenamefont{Stoffel, Rastelli,
  Tersoff, Merdzhanova, and Schmidt}}]{Rastelli2006}
\bibinfo{author}{\bibfnamefont{M.}~\bibnamefont{Stoffel}},
  \bibinfo{author}{\bibfnamefont{A.}~\bibnamefont{Rastelli}},
  \bibinfo{author}{\bibfnamefont{J.}~\bibnamefont{Tersoff}},
  \bibinfo{author}{\bibfnamefont{T.}~\bibnamefont{Merdzhanova}},
  \bibnamefont{and} \bibinfo{author}{\bibfnamefont{O.~G.}
  \bibnamefont{Schmidt}}, \bibinfo{journal}{Phy. Rev. B}
  \textbf{\bibinfo{volume}{74}}, \bibinfo{eid}{155326} (\bibinfo{year}{2006}).

\bibitem[{\citenamefont{Smith et~al.}(2003)\citenamefont{Smith, Chandrasekhar,
  Chaparro, Crozier, Drucker, Floyd, McCartney, and Zhang}}]{Drucker2003}
\bibinfo{author}{\bibfnamefont{D.}~\bibnamefont{Smith}},
  \bibinfo{author}{\bibfnamefont{D.}~\bibnamefont{Chandrasekhar}},
  \bibinfo{author}{\bibfnamefont{S.}~\bibnamefont{Chaparro}},
  \bibinfo{author}{\bibfnamefont{P.}~\bibnamefont{Crozier}},
  \bibinfo{author}{\bibfnamefont{J.}~\bibnamefont{Drucker}},
  \bibinfo{author}{\bibfnamefont{M.}~\bibnamefont{Floyd}},
  \bibinfo{author}{\bibfnamefont{M.}~\bibnamefont{McCartney}},
  \bibnamefont{and} \bibinfo{author}{\bibfnamefont{Y.}~\bibnamefont{Zhang}},
  \bibinfo{journal}{J. Cryst. Growth} \textbf{\bibinfo{volume}{259}},
  \bibinfo{pages}{232} (\bibinfo{year}{2003}).

\end{thebibliography}

\end{document}